\documentclass[11pt]{article}

\usepackage{amsmath}
\usepackage{amssymb}
\usepackage{epsfig}
\begin{document}

%%%%%%%%%%%%%%%%%%%%%%%%%%%%%%%%%%%%%%%%%%%%%%%%%%%%%%%%%%%%%%%%%%%%%

\begin{titlepage}

% the footnote symbols are only redefined for the title page !
\renewcommand{\thefootnote}{\alph{footnote}}
\vspace*{-3.cm}
\begin{flushright}

\end{flushright}

\vspace*{0.3in}

\renewcommand{\thefootnote}{\fnsymbol{footnote}}
\setcounter{footnote}{-1}

{\begin{center} {\Large\bf Perturbative Chern-Simons Theory From The
Penner Model}

\end{center}}
\renewcommand{\thefootnote}{\alph{footnote}}

\vspace*{.8cm} {\begin{center} {\large{\sc
                Noureddine~Chair
                }}
\end{center}}
\vspace*{0cm} {\it
\begin{center}
 Physics Department,
Al al-Bayt University, Mafraq, Jordan

Email: n.chair@rocketmail.com\\
\hspace{19mm}nchair@aabu.edu.jo
\end{center} }

\vspace*{1.5cm}

{\Large \bf
\begin{center} Abstract\end{center} }
\  We show explicitly that the perturbative $SU(N)$ Chern-Simons
theory arises naturally from two Penner models, with opposite
coupling constants. As a result computations in the perturbative
Chern-Simons theory are carried out using the Penner model, and it
turns out to be simpler and transparent. It is also shown that the
connected correlators of the puncture operator in the Penner model,
are related to the connected correlators of the operator that gives
the Wilson loop operator in the conjugacy class. \vspace*{.5cm}
\end{titlepage}

\renewcommand{\thefootnote}{\arabic{footnote}}
\setcounter{footnote}{0}
\newpage

%%%%%%%%%%%%%%%%%%%%%%%%%%%%%%%%%%%%%%%%%%%%%%%%%%%%%%%%%%%%%%%%%%%%%

% draft for equation labeling
%
%\globallabel{oscprob}
%\begin{align}
%\label{oscprobPCPC}
%1 \mytag \\
%\label{oscprobPCPV}
%2 \mytag
%\end{align}

\  The free energy of the Penner model \cite{penner} is the
generating function of the orbifold Euler characteristics of the
moduli space of Riemann surfaces of genus $g$, with $s$ punctures.
Computation of such a topological invariant was first computed by
Harer and Zagier \cite{zagier} by reducing a topological problem to
a combinatorial problem and then solving it.  The $SU(N)$
perturbative Chern-Simons free energy based on the $1/N$ expansion
introduced by 't Hooft \cite{'t hooft}, and  the Penner free energy
have formally similar topological expansion.  This is the case,
since both models use fatgraphs techniques that keep track of powers
of $N$. The perturbative Chern-Simons free energy \cite{vafa1} may
be written as
$$F=\sum_{g=0,h=1}C_{g,h}N^{2-2g}\lambda^{2g-2+h},$$
where $\lambda$ is the 't Hooft coupling constant and $h$ is the
number of faces (boundaries) of the triangulated Riemann surfaces.
In the Penner model $h$ is identical to the number of punctures. The
coefficient $C_{g,h}$ was shown by Witten  \cite{witten} to be
identical to  the partition function of the A-model topological open
string theory at genus $g$ with $h$ boundaries on a 6-dimensional
target space $T^{*}S^{3}$. We  will see that these coefficients
$C_{g,h}$ are related related to the orbifold  Euler characteristics
of the moduli space of Riemann surfaces of genus $g$, with $2h$
punctures.

\  The explicit expression for the Penner free energy $F = \log Z$
in terms of the of  genus and the punctures is \cite{chair1}
\begin{equation}
\label{1}
F(t,N) = \sum_{g,s} N^{2 - 2 g}(-1)^{s} t^{2 g + s-2 }
\chi_{g,s},
\end{equation}
where the coefficients $\chi_{g,s}$ are the orbifold Euler
characteristics of the moduli space of Riemann surfaces of genus $g$
with $s$ punctures; explicitly this topological invariant is given
by
\begin{equation}
\chi_{g,s} = { (-1)^{s}( 2 g - 3 + s ) ! ( 2 g - 1 ) \over ( 2 g ) !
s ! } B_{2g} \nonumber\\
\end{equation}
where $B_{2g}$ are the Bernoulli numbers. Note that the topological
expansion of the free energy used by Distler and Vafa \cite{distler}
is $F(t,N) = \sum_{g,s} N^{2 - 2 g} t^{2 -2g -s } \chi_{g,s}$. This
expansion follows from equation (\ref{1}) by simply letting
$t\rightarrow-\frac{1}{t}$.

\  Let us now consider a sum of two Penner models one with a
coupling constant $t$, and the other with coupling constant $-t$,
such that the topological expansion of the free energy in both cases
is given by equation (\ref{1}).  If the coupling constant in the
Penner model is set to be equal $\lambda/2{\pi}n$, $\lambda$ is the
Chern-Simons Coupling constant and $n$ is a positive integer.  Let
$F(\lambda,N)$ be the total free energy for the two Penner models,
then  by using equation (\ref{1}) and summing over $n$ one has
\begin{eqnarray}
\label{2}
F(\lambda,,N) &=&
\sum_{g=0}^{\infty}\sum_{n=1}^{\infty}\sum_{s=1}^{\infty} N^{2 - 2
g}(\lambda/2{\pi}n)^{2 g + s-2 } \chi_{g,s}((-1)^{s}+1)\nonumber\\
&=&2\sum_{g=0}^{\infty}\sum_{n=1}^{\infty}\sum_{p=1}^{\infty} N^{2 - 2
g}(\lambda/2{\pi}n)^{2 g + 2p-2 } \chi_{g,2p}.
\end{eqnarray}
Explicitly, if we  write $F(\lambda,N)=\sum_{g} N^{2 -
2g}F_{g}(\lambda)$, then the genus $g$ contribution $F_{g}(\lambda)$
to the free energy is nothing but the perturbative $SU(N)$
Chern-Simons free energy $F^{p}_{g}(\lambda)$ on $S^3$ \cite{vafa1},
\begin{eqnarray}
\label{3}
F_{0}(\lambda)&=&-2\sum_{n=1}^{\infty}\sum_{p=2}^{\infty}\frac{1}{
2(p-1) 2p(2p-1)}(\frac{\lambda}{2\pi{n}})^{2p-2}\nonumber\\
F_{1}(\lambda)&=& \sum_{n=1}^{\infty}
\sum_{p=1}^{\infty}\frac{B_{2}}{ {2p}}(\frac{\lambda}{2\pi{n}})^{2p}\nonumber\\
F_{g}(\lambda)&=&2\sum_{n=1}^{\infty}\sum_{p=1}^{\infty}\frac{( 2 g
- 3 + 2p ) ! ( 2 g - 1 )}{ ( 2 g ) !( 2p)!
}B_{2g}({\frac{\lambda}{2\pi{n}}})^{2g+2p-2}.
\end{eqnarray}
Note that the Bernoulli numbers $B_{g}$ in the above equation are
alternating unlike those in \cite{vafa1}, are taken to be all
positive. As one can see from the above equation, the computations
are simple and follows immediately from the Penner model.  The genus
expansion of the free energy in the perturbative Chern-Simons theory
\cite{vafa1} is obtained from the following perturbative term in
$\lambda$,
$$F^{p}(\lambda)= \sum_{j=1}^{N-1}(N-j)\sum_{n=1}ln(1-{j^2\lambda^2\over
4\pi^2n^2N^2}).$$

\  Having identified the perturbative Chern-Simons free energy
$F^{p}_{g}(\lambda)$ on $S^3$ with the extended Penner model
described above, we may use the latter to do our computations in the
peturbative Chern-Simons.  Here computations are done for both the
double-scaling limit \cite{das}, and summation over the boundaries
$p$ as well as over the integer $n$ of the peturbative Chern-Simons
\cite{vafa2}.  let us first find the continuum limit in this theory;
to do so, one needs to sum over all faces (boundaries) $p$ in the
free energy $F^{p}_{g}(\lambda)$. Using the Penner model, this
summation is equivalent to sum over all punctures. For $g=0$, Penner
model the sum over all punctures was computed explicitly
\cite{chair1}, and we obtained the following identity,
\begin{equation}
\label{4}
F_{0}(t)=-\sum_{k=1}^{\infty}\frac{1}{k(k+1)(k+2)
}t^{k}=\frac{(1-t)^{2}}{2t^{2}}\ln(1-t) -\frac{3}{4}+\frac{1}{{2t}}.
\end{equation}
Therefore, from the relation  between the peturbative
Chern-Simons and the extended Penner model summarized by equation
(\ref{2}),  the sum over boundaries for $g=0$ is,
\begin{equation}
\label{5}
F_{0}=\sum_{n=1}^{\infty}[\frac{(1-{\lambda/{2\pi{n}}})^{2}}{2(\lambda/{2\pi{n}})^{2}}
\ln(1-{\lambda/{2\pi{n}}})+\frac{(1+{\lambda/{2\pi{n}}})^{2}}{2(\lambda/{2\pi{n}})^{2}}
\ln(1+{\lambda/{2\pi{n}}})-\frac{3}{2}].
\end{equation}
The free energy is even in both $n$, and $\lambda $ as it should be,
see equation (\ref{3}). If we define a new coupling constant $\nu_n$
by $\nu_n = {2\pi N \over \lambda}[{\lambda \over 2\pi} - n]$, as in
\cite{das}, then multiplying the above sum by $N^{2}$, gives
\begin{equation}
\label{6}
N^{2} F_{0}(\lambda) =
\sum_{n=1}^\infty[{i\pi\over2}\nu_n^2+{1\over 2}\nu_n^2 \ln
(\nu_n/N) -{3\over 4}\nu_n^2 + (n \rightarrow -n)]   -N^2
\sum_{n=1}^\infty[\ln({2\pi n\over \lambda}) + {4\pi^2 n^2 \over
\lambda^2}(\ln(\lambda/2{\pi}n)({2\pi n\over \lambda})-{3\over 2})]
\end{equation}
Note the presence of the term, ${i\pi\over2}\nu_n^2$, is
responsible for the evenness of the free energy with respect to $n$
and $\lambda$. For $g=1$, The summation over boundaries are simpler
and one may use either $F_{1}(\lambda)$ given by equation (\ref{3}),
or the Penner free energy \cite{chair1}, given by
$F_{1}(\lambda)=-\frac{B_{2}}{2}\ln(1-t)$. Therefore, in this case
one has,
\begin{equation}
\label{7}
F_{1}= -{1\over 2}B_2\sum_{n=1}^\infty[i\pi+ \ln ({\nu_n
\over N}) + (n \rightarrow -n)] +B_2 \sum_{n=1}^\infty \ln ({2\pi
n\over\lambda}).
\end{equation}

Before summing over boundaries for the higher genus $g\geq2$, one
gives first summation over punctures in the Penner model. Using the
identity $-({\frac{d}{ dt}})^{2g-3}~t^{-(s+1)}= \frac{( 2 g - 3 +
s)!} {s!}~t^{2-2g-s}$, then the sum over punctures in the free
energy  $F_{g}(t) = \sum_{s}t^{2 -2g -s } \chi_{g,s}$, for $g\geq2$
is
\begin{equation}
\label{8}
F_{g}(t)=[\frac{1}{(1+t)^{2g-2}}-\frac{1}{t^{2g-2}}]\chi_{g,0},
\end{equation}
where $\chi_{g,0}=\frac{B_{2g}}{2g(2g-2)}$ is the orbifold Euler
characteristic of the moduli space without punctures. Incidently,
the topological expansion for $F_{g}(t)$ studied in \cite{distler}
has two critical points, namely $t=-1$ and $t=0$. As we made it
clear in this paper, the peturbative Chern-Simons is connected to
the original topological expansion of the Penner model studied in
\cite{chair1},i.e, equivalent to  letting  $t\rightarrow
-\frac{1}{t}$ in equation (\ref{8}). Therefore the sum over the
punctures in the original Penner model reads
\begin{equation}
\label{9} F_{g}(t) =[{(1-1/t)^{2-2g}}-{t^{2g-2}}]\chi_{g,0}.
\end{equation}
Now, as we did for for $g=0$ and $g=1$, the sum over the boundaries
in the perturbative Chern-Simons for $ F_{g}(\lambda)$ is the sum
over $n$ of
$F_{g}(\frac{\lambda}{2\pi{n}})+F_{g}(-\frac{\lambda}{2\pi{n}})$,
that is,
\begin{equation}
\label{10}
F_{g}(\lambda) = \chi_{g,0} \sum_{n=1}^\infty[ \nu_n^{2-2g} + (n
\rightarrow -n)] - 2 \chi_{g,0} ({\lambda \over 2\pi N})^{2g-2}
~\zeta (2g-2),
\end{equation}
where $\zeta (2g-2)=\sum_{n=1}^\infty\frac{1}{n^{2g-2}} $. The
Chern-Simons theory coupling constant $\lambda$ is related to the
level of the Kac-Moody algebra $k$ by $\lambda = {2\pi N \over
k+N}$; this shows that $\lambda$ has a fundamental domain between
$0$ and $2\pi$. Therefore, the natural critical double scaling limit
would be, $$ \lambda \rightarrow 2\pi~~~~~~~~~\nu_1 = {\rm
finite}.$$ Note that at this  critical point
$\lambda/2\pi\rightarrow 1$ that is, $t\rightarrow1$ which is
nothing  but the  critical point in the Penner model \cite{chair1}.

\  We turn now to the sum over the boundaries $p$, and over $n$ of
the free energy $F_{g}(\lambda)$; this sum is known to be connected
to the Hodge integrals and Gromov-Witten Theory \cite{faber}. For
$g=0$, the computations are easy and straightforward.  From equation
(\ref{5}) one has
\begin{equation}
\label{11}
(\frac{N}{\lambda})^{2}F_{0}(\lambda)
=(\frac{N}{\lambda})^{2}\sum_{n=1}^{\infty}[4\pi^{2}n^{2}
\frac{(1-{\lambda/{2\pi{n}}})^{2}}{2}
\ln(1-{\lambda/{2\pi{n}}})+4\pi^{2}n^{2}\frac{(1+{\lambda/{2\pi{n}}})^{2}}{2}
\ln(1+{\lambda/{2\pi{n}}})-\frac{3\lambda^{2}}{2}].
\end{equation}
Differentiating $F_{0}(\lambda)$, with respect to $\lambda$, twice
gives $d^{2}/ {d{\lambda}^{2}}F_{0}(\lambda)=\sum_{n\in {\bf
Z}}^{'}[\ln(1-{\lambda/{2\pi{n}}})]$, where $n=0$ is not included in
the sum. From the product formula ${sin(\pi x)\over \pi
x}=\prod_{n=1}(1-{x^2\over n^2})$, one has the identity
\begin{equation}
\label{12}
\sum_{n\in {\bf Z}, n\neq 0} [\ln(1-{\lambda/{2\pi{n}}})]=
i{\lambda}/2+\ln(1-e^{-i\lambda})-\ln{\lambda}-i\frac{\pi}{2}.
\end{equation}
Therefore, we see that our computations using the Penner model are
simpler and transparent.  The summed free energy $F_{0}(\lambda)$ is
obtained simply by integrating twice the expression for $d^{2}/
{d{\xi}^{2}}F_{0}(\xi)$ with respect to $\xi$ from $0$ to $\lambda$.
Note that, here we follow closely the same lines in deriving the
product formula for $\sin{x}$. Taking $\ln{\lambda}= 2i\pi{m}$,
$\theta=0$, the cut line being the real axis then
\begin{equation}
\label{13}
F_{0}(\lambda)=\zeta(3)+i\zeta(2)-i(m+1/4)\pi\lambda^{2}
+i\lambda^{3}/12+\sum_{n=1}^{{\infty}}\frac{e^{-in\lambda}}{n^{3}},
\end{equation}
this is identical to the result obtained in \cite{vafa2}. The
coefficient of the last term in the above equation is the $g=0$
Gromov-Witten invariant of a Calabi-Yau 3-fold, $C(0,n)=
\frac{1}{n^{3}}$ \cite{faber}. For $g=1$ one can see that the free
energy up to constat terms is given by
\begin{equation}
\label{14}
F_{1}(\lambda)=\sum_{n\in {\bf Z}, n\neq 0}
[\ln(1-{\lambda/{2\pi{n}}})]=-B_{2}/2[i{\lambda}/2+\ln(1-e^{-i\lambda})].
\end{equation}
The Gromov-Witten invariant in this case  follows from the second
term and is given by $C(1,n)= \frac{1}{12{n}}$.

\  Now, we come to the sum over boundaries $p$ for $g\geq2$, as we
explained before this is equivalent to sum over punctures in the proposed
penner model.  By rewriting equation (\ref{10}), explicitly in terms
of $\lambda$, then we have
\begin{equation}
\label{15}
F_{g}(\lambda) =
({\frac{N}{\lambda}})^{2-2g}\chi_{g,0}\sum_{n\in {\bf Z}, n\neq 0}[
(\lambda-2\pi{ n})^{2-2g}] - 2({\frac{N}{\lambda}})^{2-2g}
\chi_{g,0} ({1 \over 2\pi})^{2g-2} ~\zeta (2g-2),
\end{equation}
the  sum over $n$ may be carried out by simply using the product
formula for $\sin\pi{x}$, from which we obtain $\sum_{n\in {\bf Z}}
\ln(\lambda-2\pi{ n}) \approx\ln(1-e^{-i\lambda})$. Note that the
terms that are not written would disappear upon differentiating
$(2{g}-2)$ times the right hand side of the approximation, and so we
have
\begin{equation}
\label{16}
F_{g}(\lambda)=
({\frac{N}{\lambda}})^{2-2g}(-1)^{g-1}\chi_{g,0}
\frac{1}{(2g-3)!}\sum_{n\geq1}{n^{2g-3}}{e^{-in\lambda}} -
2({\frac{N}{\lambda}})^{2-2g} \chi_{g,0} ({1 \over 2\pi})^{2g-2}
~\zeta (2g-2).
\end{equation}
The coefficient of the first term may be written as $|\chi_{g,0}|
\frac{n^{2g-3}}{(2g-3)!}$, which is nothing but the Gromov-Witten
invariant $C(g,n)$ \cite{faber}. The coefficient of the second term
$2\chi_{g,0} ({1 \over 2\pi})^{2g-2} ~\zeta (2g-2)$ is the degree
zero Gromov-Witten invariant \cite{vafa2}.  This is identical to the
Hodge integral, $\int_{{\cal M}_g} c_{g-1}^3$, \cite{faber},
\cite{vafa1} where $c_{g-1}$ is the $(g-1)$ Chern class of the Hodge
bundle.

\  We now push further the connection between the Penner model and
the perturbative Chern-Simons theory. We carry out this connection
by considering correlators in the Penner model that exhibit
logarithmic singularities, and find out the corresponding
correlators in the C.S theory. In the former the correlators that
exhibit logarithmic scaling violation are the puncture operators
$\partial/\partial{\mu}$ \cite{distler} \cite{chair1} \cite{chair2},
where $\mu=N(1-t)$. The free energy $F_{g}(\mu)$ is known to be
related to the orbifold Euler characteristics $\chi_{g,0}$. When
differentiated $s$-times, then we obtain a generating function for
the orbifold Euler characteristics with $s$ punctures.  This
procedure is equivalent to put $s$ punctures on a Riemann surface.
We have shown in \cite{chair2} that the puncture operator
$\partial/\partial{\mu}$ is identified with the operator ${\rm
Tr}\ln(1-i\surd{t}M)$, where $M$ is an $N\times N$ hermitian matrix.
The equivalence of the two operators, was checked for the one point
and the two point connected correlators.  Explicitly the $k^{th}$
power for the operator ${\rm Tr}\ln(1-M)$ reads,
\begin{eqnarray}
\label{17} \frac{1}{k!}({\rm Tr}\ln (1-M))^{k}&=&\frac{1}{k!}
(-1)^{k}(\sum_{j\geq 1}{\rm Tr}\frac{M^{j}}{j})^{k}\nonumber\\
&=&(-1)^{k}\frac{1}{k!}\sum_{k_{1}\geq 0,k_{2}\geq
0,\cdots}\frac{k!}{k_{1}!k_{2}!\cdots} (\frac{{\rm
Tr}{M}}{1})^{k_{1}}(\frac{{\rm Tr}{M^{2}}}{2})^{k_{2}}
\cdots  \nonumber\\
&=&(-1)^{k}\sum_{k_{1}\geq 0, k_{2}\geq
0,\cdots}\frac{1}{\prod_{j\geq 1} k_{j}!j^{k_{j}}}~
\prod_{j=1}^{\infty}({\rm Tr}{M^{j}})^{k_{j}},
\end{eqnarray}
where the sum is taken over all  $k's$,  zero or positive integers such that
$\sum k_{j}=|\vec{k}|=k$. Therefore, the connected $k^{th}$ correlators of the puncture
operator may be be written as
\begin{eqnarray}
\label{18}
\frac{1}{k!}\langle({\rm Tr}\ln (1-M))^{k}\rangle^{c}&=&
(-1)^{k}\sum_{k_{1}\geq 0, k_{2}\geq
0,\cdots}\frac{1}{z_{\vec{k}}}~\langle\Upsilon_{\vec{k}}(M)\rangle^{c},
\end{eqnarray}
where, $z_{\vec{k}}=\prod_{j\geq 1}~ k_{j}!j^{k_{j}}$,
$\Upsilon_{\vec{k}}(M)=\prod_{j=1}^{\infty}({\rm Tr}M^{j})^{k_{j}}$
this formally looks like the operator that gives the Wilson loop
operator in the conjugacy class basis \cite{marino}.  Therefor the
correlators of the puncture operator in the Penner model, is very
close to the logarithm of the expectation of the Ooguri-Vafa
operator \cite{vafa3}, \cite{marino} $\ln\langle
Z(U,V)\rangle=\sum_{\vec{k}}\frac{1}{z_{\vec{k}}}~
\langle\Upsilon_{\vec{k}}(U)\rangle^{c}~\Upsilon_{\vec{k}}(V)$. When
the $M\times M$ matrix  $V$ (source term), is the unit matrix then
formally  correlators of the puncture operator and the above
generating function are identical up to  a constant factor.
Therefore, it is possible to compute
$\langle\Upsilon_{\vec{k}}(U)\rangle^{c}$ from the connected
correlators of the Penner model \cite{chair1}.  These correlators
are written in terms of the orbifold Euler characteristics with
punctures.

\ Finally we point out that the connection between the perturbative Chern-Simons theory, and the Penner model, may also be seen, from the explicit expression for
the free energy of the Penner model \cite{chair1},
$$F(t,N)=\sum_{m=1}^{\infty}\frac{B_{2m}}{
2m(2m-1)}(\frac{t}{N})^{2m-1}+
\sum_{p=1}^{N-1}(N-p)\ln(1-\frac{pt}{N}).$$  To obtain the
perturbative Chern-Simons  free energy, we follow the same procedure
used in the paper. Let $t\rightarrow-t$, then one has
$$F(t,N)+F(-t,N)=
\sum_{p=1}^{N-1}(N-p)\ln(1-\frac{p^{2}t^{2}}{N^{2}}).$$ Next, set
$t=\frac{\lambda}{2{\pi}n}$, then summing over $n\geq1$ gives the
perturbative Chern-Simons free energy \cite{vafa1}.  It remains to
see the Physical justification and interpretation  of the connection
between the perturbative Chern-Simons theory and the extended Penner
model, proposed in this paper.

\vspace{7mm} {\bf Acknowledgment:} I would like to thank M.Marino
for reading and making comments on this work .  I would also like to
thank G.Bonelli, L.Bonora, K.S.Narain, and E.H.Saidi for discussions
during the early stages of this work, H.M Al-Nasser and M.I Serhan
for their useful remarks, Al al-Bayt university, the Abdus Salam
center ICTP, for the support.

\newpage

\bibliographystyle{phaip}

\begin{thebibliography}{1}
\bibitem{penner}
R.C.Penner,J.Diff.Geom. \textbf{27} (1998) 35
\bibitem{zagier}
J.Harer and D.Zagier, Inv.Math.\textbf{85} (1986) 457
\bibitem{'t hooft}
G.'t Hooft, Nucl.Phys. B.\textbf{72} (1974) 461
\bibitem{vafa1}
R.Gopakumar and C.Vafa, "M-theory and topological strings,I" e-print
hep-th/9812127
\bibitem{witten}
E.Witten, Prog.Math. \textbf{133} (1995) 637
\bibitem{chair1}
N.Chair, Rev.Math.Phys. \textbf{3} (1991) 285
\bibitem{distler}
J.Distler and C.Vafa, Mod.Phys.Lett. \textbf{A6} (1991) 259
\bibitem{das}
Sumit R.Das and C.Gomez, JHEP. \textbf{0410} (2004) 026
\bibitem{vafa2}
R.Gopakumar and C.Vafa, Adv.Theor.Math.Phys. \textbf{3} (1999) 1415
\bibitem{faber}
C.Faber and R.Pandharipande, Invent.Math. \textbf{139} (2000) 173
\bibitem{chair2}
N.Chair and S.Panda, Phys.Lett. \textbf{272B} (1991) 230
\bibitem{marino}
M.Marino, Rev.Mod.Phys. \textbf{77} (2005) 675
\bibitem{vafa3}
H.Ooguri and C.Vafa, Nucl.Phys. B \textbf{577} (2000) 419.

\end{thebibliography}

\end{document}